\documentclass[aps, prx, reprint, showpacs, longbibliography,
               superscriptaddress]{revtex4-1}

\usepackage{mathtools}
\usepackage{siunitx}
\usepackage{dcolumn}

\usepackage{paralist}
\usepackage{verbatim}
\usepackage{afterpage}
\usepackage{siunitx}
\usepackage[inline]{enumitem}

\usepackage[utf8]{inputenc}
\usepackage{microtype}
\usepackage{txfonts}
\usepackage[upright]{txgreeks}
\usepackage[varl]{inconsolata}
\usepackage[mathcal]{euscript}
\usepackage[usenames, dvipsnames, svgnames, table]{xcolor}
\usepackage[colorlinks=true,
            citecolor=MidnightBlue,
            linkcolor=Maroon,
            urlcolor=DarkGreen]{hyperref}

\usepackage{graphicx,calc}
\usepackage[format=plain, justification=centerlast, small]{caption}
\usepackage{subcaption}
\usepackage{tikz}
\usetikzlibrary{arrows,
                petri,
                topaths}

\newcolumntype{d}[1]{D{.}{.}{#1}}
\newcolumntype{G}{>{\centering\arraybackslash}m{5em}}

\DeclareSIUnit\eV{eV}
\DeclareSIUnit\parsec{pc}
\DeclareSIUnit\yr{yr}

\newcommand{\km}{\kilo\meter}

\newcommand{\Hz}{\hertz}
\newcommand{\kHz}{\kilo\hertz}

\setlist[enumerate]{nosep}

\begin{document}

\newcommand{\LIGOMIT}{LIGO Laboratory, Massachusetts Institute of Technology, Cambridge, MA 02139, USA}
\newcommand{\LIGOCaltech}{LIGO Laboratory, California Institute of Technology, Pasadena, CA 91125, USA}

\title{Metrics for next-generation gravitational-wave detectors}
\author{Evan D. Hall}
\affiliation{\LIGOMIT}
\author{Matthew Evans}
\affiliation{\LIGOMIT}

\begin{abstract}
    Gravitational-wave astrophysics has the potential to be transformed by a global network of longer, colder, and thus more sensitive detectors.
    This network must be constructed to address a wide range of science goals, involving binary coalescence signals as well as signals from other, potentially unknown, sources.
    It is crucial to understand which network configurations---the number, type, and location of the detectors in the network---can best achieve these goals.
    In this work we examine a large number of possible three-detector networks, variously composed of Voyager, Einstein Telescope, and Cosmic Explorer detectors, and evaluate their performance against a number of figures of merit meant to capture a variety of future science goals.
    From this we infer that network performance, including sky localization, is determined most strongly by the type of detectors contained in the network, rather than the location and orientation of the facilities.
\end{abstract}

\maketitle

\section{Introduction}
\label{sec:introduction}

Gravitational-wave science, which started in earnest with the first detection of gravitational-waves in 2015~\cite{Abbott:2016blz}, has immense unexplored potential.
Aiming to exploit this potential, the gravitational-wave community is already considering designs for next generation of detectors~\cite{ETBook,Evans:2016mbw}.
It is clear that to get the most science out of the gravitational-wave signals we detect, a network of large-scale observatories will be required.
However, at present there are many unanswered questions about how best to construct such a network.

In this paper we will address several of the fundamental questions required to maximize the scientific potential of a terrestrial gravitational-wave detector network.
These questions include:
Which science goals are sensitive to the location and orientation of the detectors in the network, and in what way?
How the does design of the network's constituents impact its output?
What science can be done with a heterogeneous mix of second and third generation detectors?
 
\begin{figure}[t]
    \includegraphics[width=\columnwidth]{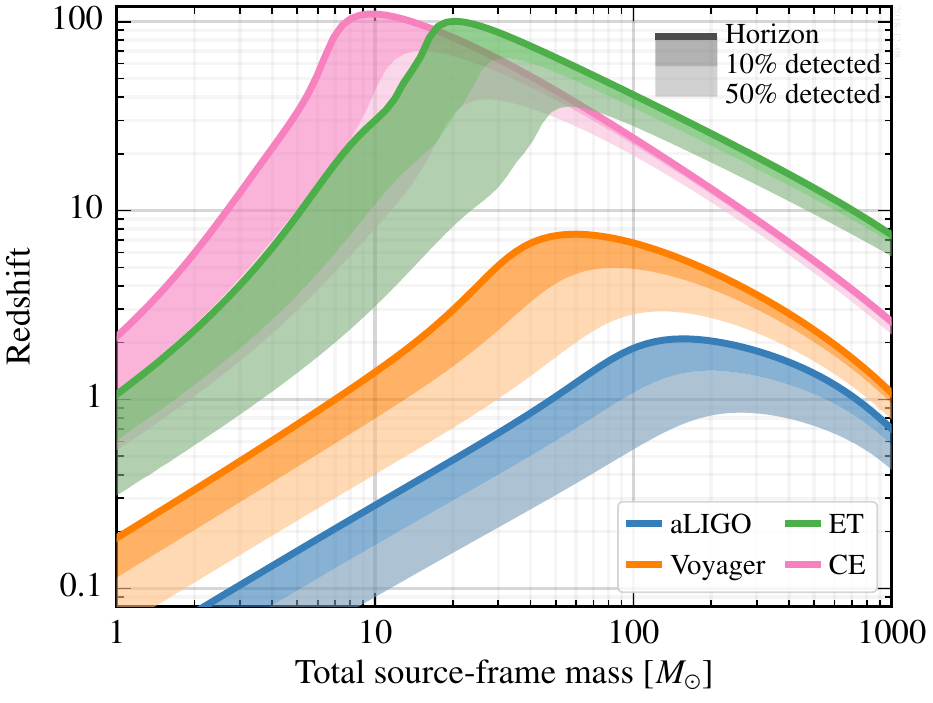}
    \caption{
        The response distance~\cite{T1500491,Chen:2017wpg} of selected second- and third-generation detectors for equal-mass, nonspinning binaries, shown as a function of total source-frame mass.
        The binaries are distributed isotropically in sky location and inclination angle.
        The solid lines denote the horizon---the redshift beyond which none of the sources are detected.
        The shaded bands then show the redshifts at which 10\% and 50\% of the sources at that redshift are detected.
        Here a source is assumed to be detected if it appears in a detector with matched-filter signal-to-noise ratio (SNR) $\ge 8$.
        The detectors considered here are Advanced LIGO (aLIGO)~\cite{Harry:2010zz}, Voyager~\cite{VoyagerWP}, Einstein Telescope (ET)~\cite{ETBook}, and Cosmic Explorer (CE)~\cite{Evans:2016mbw}.
        The response distance is a measure of detector sensitivity only; it does not include assumptions about the redshift distribution of astrophysical sources.
        }
    \label{fig:gw_horizons}
\end{figure}

\begin{figure}[t]
    \includegraphics[width=\columnwidth]{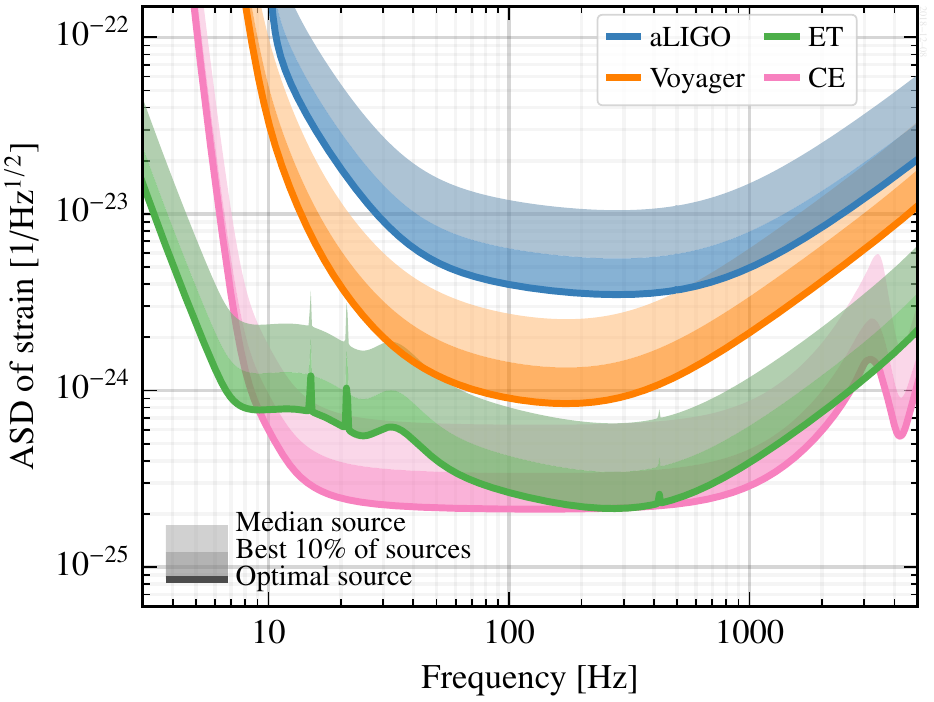}
    \caption{
        Effective strain amplitude spectral density (ASD) of selected second- and third-generation detectors for monochromatic sources distributed isotropically in sky position, inclination, and polarization.
        The solid lines denote the effective strain sensitivity for an optimally oriented source.
        The bands then denote the effective sensitivity for the best 10\% of sources, and the median source.
        The detectors considered here are Advanced LIGO (aLIGO), Voyager, Einstein Telescope (ET), and Cosmic Explorer (CE).
        }
    \label{fig:strains}
\end{figure}

Instrument and facility designs often focus on optimizing the performance of single detectors rather than networks.
Such optimizations have relied on metrics such as the inspiral range---the distance out to which each detector could detect a model system (usually a 1.4--1.4\,$M_\odot$ binary coalescence) with a certain signal-to-noise ratio~\cite{Finn:1992xs}---or the detector's strain sensitivity~\cite{Miao:2013hga}.
A variety of cosmological generalizations of these range metrics have been developed~\cite{Chen:2017wpg}; one such generalization, the response distance, is shown in Fig.~\ref{fig:gw_horizons}.
Effective strain sensitivities are shown in Fig.~\ref{fig:strains}, taking into account effects from frequency-dependent antenna patterns~\cite{Essick:2017wyl} and multi-interferometer detectors.

Systematic studies of gravitational-wave detector network optimization, measured quantitatively via a set of metrics, go back more than a decade.
Searle et al.~\cite{Searle:2001hz} considered how coincident detections of binary neutron stars could be augmented by the addition of another facility to the the existing gravitational-wave network.
Schutz~\cite{Schutz:2011tw} proposed three figures of merit for a gravitational-wave network: the triple-coincidence detection rate, the isotropy of detections across the sky, and the typical localization of events on the sky.
Raffai et~al.~\cite{Raffai:2013yt} optimized the facility placement of a set of triangular (Einstein-Telescope-like) detectors as well as the placement of a LIGO facility in India;
this procedure was then generalized by Hu et~al.~\cite{Hu:2014bpa}.
Michimura et~al.~\cite{Michimura:2018sih} optimized the optical configuration of the Kagra detector~\cite{Aso:2013eba} with respect to the sky localization performance of the global advanced detector network.

More generally, others~\cite{Mills:2017urp,Vitale:2016icu,Vitale:2018nif} have already examined the performance of a set of plausible third-generation gravitational-wave networks against some set of metrics.
The binary-neutron-star (BNS) localization capabilities of networks with third-generation detectors has been explored by Mills et~al.~\cite{Mills:2017urp}.
The binary-black-hole (BBH) parameter estimation capabilities---including masses, spins, redshift, and localization---of networks with third-generation detectors has been explored by Vitale and Evans~\cite{Vitale:2016icu} and Vitale and Whittle~\cite{Vitale:2018nif}.

While these works evaluate the performance of some networks, and optimize a few assemblies of detectors for some performance metrics, they do not address the critical questions posed earlier in this section.
In this work, we present in full the performance of a large ensemble of networks against a list of metrics, in order to see the full landscape of network performance.
This reveals which metrics require careful optimization, and which are relatively insensitive to network configuration.

\section{Metrics that connect network parameters with science goals}
\label{sec:goals-metrics-networks}

The questions raised in Sec.~\ref{sec:introduction} naturally drive one to ``optimally'' choose the network parameters---the number, type, location, and orientation of third-generation instruments across the globe---based on what information is needed to address specific science goals.
These goals are quite diverse; e.g., determining the history of star formation, testing corrections to general relativity, uncovering the nuclear physics inside neutron stars and supernovae, and constraining the dark energy equation of state.
Each of these may require different kinds of observational information or levels of precision.
In addition, each of these science goals involves a variety of data analysis techniques and scientific products whose connection to the network parameters is not immediately obvious.

For these reasons, network optimization studies~\cite{Hu:2014bpa,Raffai:2013yt,Michimura:2018sih} often focus on optimizing a smaller number of metrics, by which we mean intermediate data products that have a reasonably clear dependence on the network parameters and which then feed into the more specialized analysis that is used to address the science goals.
Such metrics include signal-to-noise ratio, polarization sensitivity, and source localization.

In addition to choosing a set of metrics, one must also consider whether to optimize these metrics for the \emph{best} events, or the \emph{typical} events.
Certain studies benefit from the large statistics of the total population of events, thereby suggesting that metrics should be optimized for the median event.
Other studies benefit from collecting a few of the loudest events, thereby suggesting that metrics should be optimized for loudest events.
Still other studies require collecting a few rare or special events, which may not be captured in the generic metrics; these studies may need to optimize the median event so as not to lose out on the rare events.

In the following section we choose a few example metrics and explore how different third-generation networks perform.
Where appropriate, we examine how networks perform for both the average (median) event and the best 10\% of events.

\section{Metrics}

\begin{table}[t]
    \centering
        \begin{ruledtabular}
        \begin{tabular}{l l *{3}{D..{-1} }}
            Code  & Location           & \text{Lat.}   & \text{Long.} & \theta_\text{XE} \\
            \hline
            H     & Hanford, USA       &  46.5 & -119.4 &  126 \\
            L     & Livingston, USA    &  30.6 &  -90.8 & -162 \\
            V     & Pisa, Italy        &  43.6 &   10.5 &   71 \\
            I*    & India              &  14.2 &   76.4 &   45 \\
            K     & Kamioka, Japan     &  36.4 &  137.3 &   28 \\
            E*    & Europe             &  47.4 &    8.5 &   11 \\
            A*    & Western Australia  & -31.5 &  118.0 &  -58 \\
            U*    & Utah, USA          &  40.8 & -113.8 &  -30 \\
        \end{tabular}
        \end{ruledtabular}
    \caption{
        Coordinates and orientations for the facilities considered in this work.
        $\theta_\text{XE}$ is the counterclockwise angle from due east made by the X-arm (or for ET, by the base of the triangle).
        Facilities with an asterisk have not been constructed and the coordinates are chosen for illustrative purposes only.
        Facility coordinates are rounded to the nearest $0.1^\circ$ (${\sim}\SI{10}{\km}$ accuracy).
        }
    \label{tab:facility_coordinates}
\end{table}

We have chosen several metrics relating to parameter estimation for coalescing binaries, as well as metrics intended to evaluate network performance for neutron-star/supernova physics and stochastic backgrounds:
\begin{enumerate}
    \item the sky localization area of 1.4--1.4\,$M_\odot$ binary neutron star coalescences at redshift $z = 0.3$, a metric useful for assessing the feasibility of electromagnetic followup;
    \item \label{itm:bbh_snr_metric} the signal-to-noise ratio of 30--30\,$M_\odot$ binary black hole coalescences at redshift $z = 2$, a metric for quantifying the quality of the most frequent events;
    \item \label{itm:bbh_dist_metric} the distance uncertainty (or equivalently, redshift uncertainty) of the 30--30\,$M_\odot$ binary black hole coalescences at $z = 2$, a metric useful for assessing the constraining power of third-generation networks on the stellar evolution history of the universe;
    \item \label{itm:bbh_inc_metric} the inclination angle uncertainty of the same 30--30\,$M_\odot$ systems at $z = 2$,
    which is representative of the network's ability to distinguish between gravitational-wave polarizations;
    \item the signal-to-noise ratio of a high-frequency strain signal, a metric useful for studies of the post-merger signal in neutron star coalescences, and for galactic supernovae; and
    \item the level of stochastic background that could be observed with $\text{SNR} = 1$ after one year of observation time.
\end{enumerate}
Note that while these metrics have chosen to be representative of the basic kinds of signal searches and parameter estimation problems that are carried out with gravitational-wave detectors, they feed more broadly into overarching science goals.
For example, for tests of general relativity, the SNR performance for binary black-hole systems as quantified by metric~(\ref{itm:bbh_snr_metric}) is related to the precision with which one can perform spectroscopy on black hole ringdown signals~\cite{Berti:2016lat}, and the metric~(\ref{itm:bbh_inc_metric}) is related to polarization-discrimination ability required to carry out a broad class of polarization-based relativity tests~\cite{Will:2014kxa}.

Our strategy for exploring the above metrics is as follows.
First, we write down a list of possible three-facility configurations using various combinations of second- and third-generation detectors~\footnote{Note the distinction between \emph{detector} and \emph{facility}: while each Cosmic Explorer facility is assumed to contain one Cosmic Explorer detector, each Einstein Telescope facility contains \emph{three} Einstein Telescope detectors, arranged in an equilateral triangle.}; in the end we select nine such combinations.
For each of these nine configurations, we generate an ensemble of possible networks by allowing the locations of the third-generation facilities (ET or CE) to vary randomly across the globe~\footnote{We do not take into account likely geographical constraints such as oceans.
This results in some highly improbable networks, but it means that our exploration of network topology is exhaustive: one should \emph{not} expect to find a geographically forbidden network that vastly outperforms the geographically realistic ones.}, and by allowing the Voyager detectors to be chosen randomly from one of five facilities (Hanford, Livingston, Pisa, India, and Kamioka).
For each of the configurations involving at least one ET or CE facility, we generate 200 random networks; for the three-Voyager configuration we generate 10 networks (the maximum possible by choosing 3 facilities from a set of 5); and for the HLV configuration we use only the one existing network with two LIGO detectors at Hanford and Livingston and one Virgo detector~\cite{TheVirgo:2014hva} at Pisa.
Any network realization with the facilities placed too close together (area of the planar triangle spanned by the facilities is less than $0.25{r_\oplus}^{\!\!2}$, where $r_\oplus$ is the radius of the Earth) is rejected.
In total, this results in more than 1000 networks whose performance we evaluate.
In the subsequent plots, the random networks are shown as circles, with the color of each circle corresponding to the nine configurations given above.

For each of the nine configurations we generate a single, plausible network using the facility coordinates given in Tab.~\ref{tab:facility_coordinates}, with detectors assigned to facilities according to Tab.~\ref{tab:fixed_network_configurations}.
These are shown as colored stars in the subsequent plots.

For the metrics involving binary coalescences, we generate $12\times8^2 = 768$ systems distributed isotropically in the sky (via the HEALPix scheme~\cite{Gorski:2004by}\footnote{\url{https://healpix.sourceforge.io}}), and with random inclinations.
The components of the binary are non-spinning and have equal mass.
For each of the ${>}1000$ networks, we evaluate each metric against all of these systems, resulting in a distribution of 768 values; we then extract the median value of this distribution, and the value delimiting the best 10\% of the distribution.
No SNR cuts or trigger thresholds are applied.

The stochastic metric results in only one value per network, so no computation of quantiles is necessary.

\begin{table}[t]
    \begin{ruledtabular}
        \begin{tabular}{l c c c c c c c c c}
            Network     & H     & L     & V     & I     & K     & E     & A     & U     \\
            \hline
            HLV         & aL    & aL    & AdV   & ---   & ---   & ---   & ---   & ---   \\
            3Voy        & Voy   & ---   & Voy   & Voy   & ---   & ---   & ---   & ---   \\ 
            ET/2Voy     & ---   & Voy   & ---   & ---   & Voy   & ET    & ---   & ---   \\
            CE/2Voy     & ---   & ---   & Voy   & ---   & Voy   & ---   & ---   & CE    \\
            ET/CE/Voy   & ---   & ---   & ---   & Voy   & ---   & ET    & ---   & CE    \\
            2CE/Voy     & ---   & ---   & ---   & ---   & Voy   & ---   & CE    & CE    \\
            ET/2CE      & ---   & ---   & ---   & ---   & ---   & ET    & CE    & CE    \\
            3CE         & ---   & ---   & ---   & ---   & ---   & CE    & CE    & CE    \\
            3ET         & ---   & ---   & ---   & ---   & ---   & ET    & ET    & ET    \\
        \end{tabular}
    \end{ruledtabular}
    \caption{Composition of plausible network configurations, shown as stars in the subsequent plots. ``aL'' is Advanced LIGO, ``AdV'' is Advanced Virgo; ``Voy'' is LIGO Voyager; ``ET'' is Einstein Telescope, and ``CE'' is Cosmic Explorer.
    H, L, V, I, K, E, A, and U are facilities whose coordinates are given in Tab.~\ref{tab:facility_coordinates}.}
    \label{tab:fixed_network_configurations}
\end{table}

\subsection{Localization of neutron-star binaries at $z = 0.3$}

\begin{figure}[tp]
        \includegraphics[width=\columnwidth]{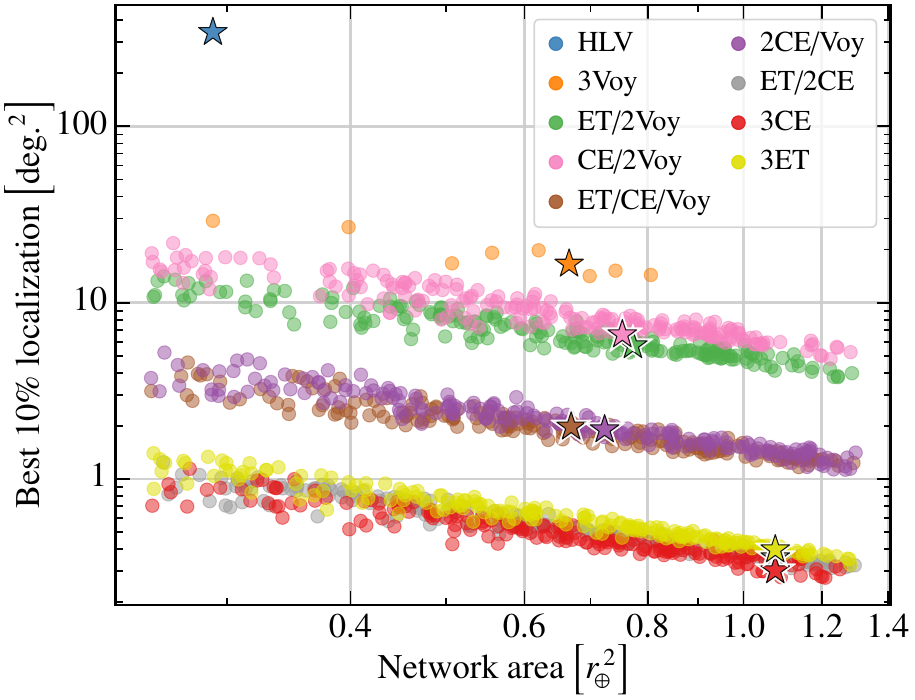}
    \caption{
        Distribution of best 10\% sky localization areas (90\% confidence) for randomly simulated networks, shown for $z = 0.3$ and $M_1 = M_2 = 1.4\,M_\odot$.
        The horizontal axis denotes the area of the triangle spanned by the three facilities in units of earth radii squared.
        Stars indicate plausible network configurations (Tab.~\ref{tab:fixed_network_configurations}).
        }
    \label{fig:cbc_loc_bns}
\end{figure}

Out to redshift $z = 0.3$, one may reasonably expect ${>}500$ binary neutron star coalescence events to pass through the earth every year, even assuming a pessimistic local merger rate density ${\sim}\SI{100}{Gpc^{-3}\,yr^{-1}}$.
These events, if sufficiently localized, will be within the followup capabilities of next-generation telescopes.

With a network of three separate detector facilities, events can be localized to an ellipse on the sky.
In this work we compute the localization area (90\% confidence interval) via the basic Fisher matrix procedure described by Singer and Price~\cite[\S B]{Singer:2015ema}, with uniform priors.

In Fig.~\ref{fig:cbc_loc_bns} we plot the resulting distributions for the best 10\% of localizations, with the networks sorted by the area spanned by the three facilities.
As expected, the localization performance scales inversely with the area.
It is also clear that each addition of a 3G facility to the network significantly improves the network's localization capability.
A network composed of two 2G facilities and one 3G facility is roughly a factor of 3 better than a baseline set of three 2G facilities.
Including two 3G facilities offers an order of magnitude improvement over the baseline, while a full 3G network offers another factor of 4 or 5, with a significant fraction of localizations below \SI{1}{deg.^2}.

\begin{figure}[tp]
    \includegraphics[width=\columnwidth]{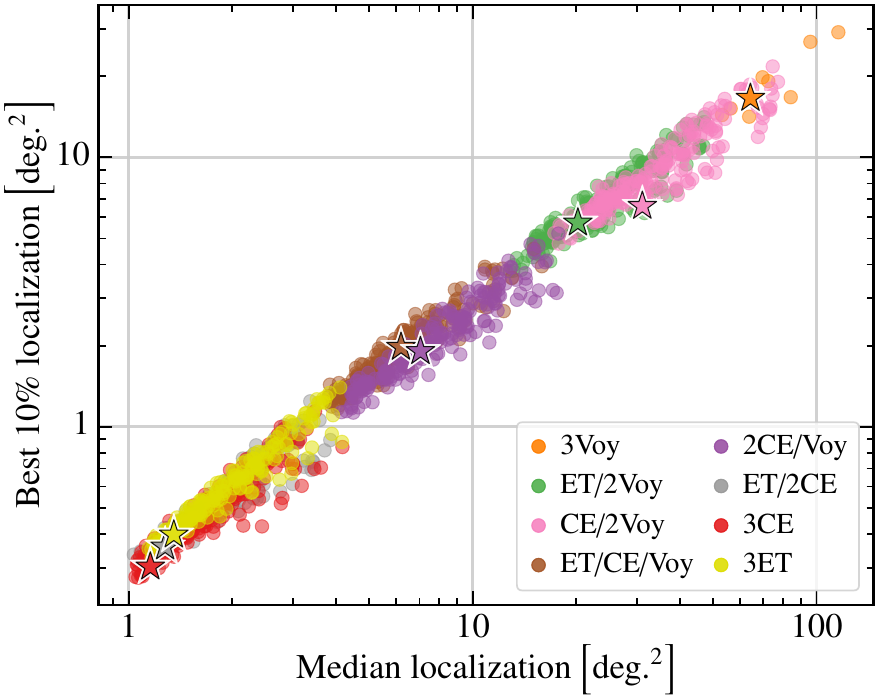}
    \caption{Scatter plot showing median sky localization and best 10\% sky localization (90\% confidence) for $z = 0.3$ and $M_1 = M_2 = 1.4\,M_\odot$.
        The HLV point is omitted.
        Stars indicate plausible facility locations, as given in Tab.~\ref{tab:fixed_network_configurations}.}
    \label{fig:cbc_med_better_loc_bns}
\end{figure}

In Fig.~\ref{fig:cbc_med_better_loc_bns} we plot the best 10\% of localizations against the median localizations.
This result shows that these two metrics are highly correlated, indicating that there is no need to trade high performance on the best events for good performance on the majority of events, at least for localization capability.

\subsection{Signal-to-noise ratios of black-hole binaries at $z = 2$}

\begin{figure}[tp]
    \includegraphics[width=\columnwidth]{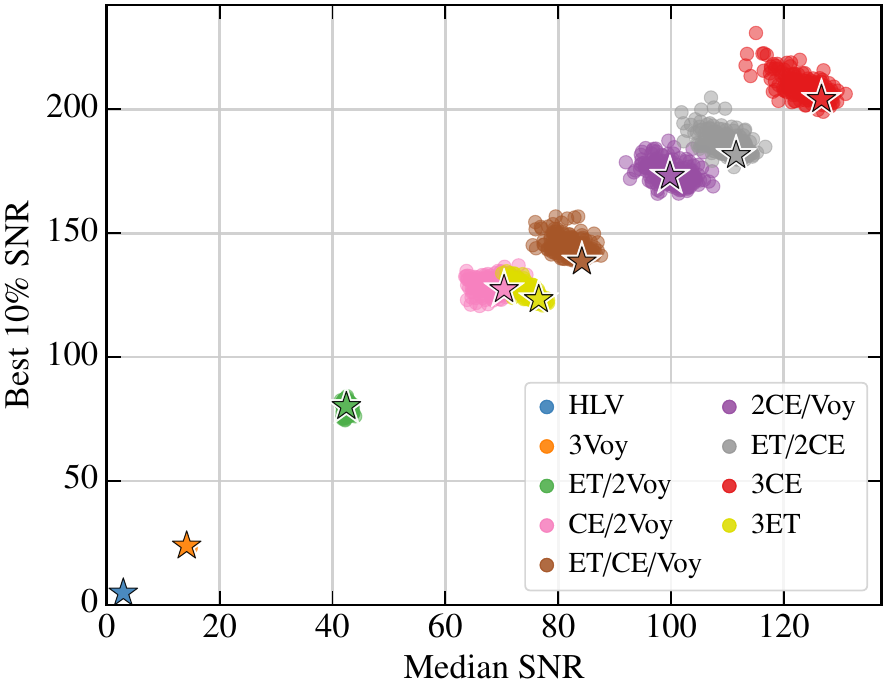}
    \caption{Scatter plot showing median SNR and best 10\% SNR for a $M_1 = M_2 = 30\,M_\odot$ binary black-hole coalescence at redshift $z = 2$.}
    \label{fig:cbc_med_better_snr_bbh}
\end{figure}

To evaluate the signal-to-noise ratio (SNR) performance of our ensemble of networks, we compute the matched-filter network SNRs for a population of 30--30\,$M_\odot$ binary black hole coalescences at $z = 2$ distributed isotropically in the sky and with random inclinations.
Fig.~\ref{fig:cbc_med_better_snr_bbh} shows the performance of the ensemble of networks by plotting each network's median SNR against its best 10\% SNR.
Unlike the distribution of localizations, the distribution of SNRs has comparatively little dependence on the location and orientation of the detectors; instead, the network performance is determined predominantly by the network composition, with a scatter ${\lesssim}30\%$.

For the networks consisting of two or more third-generation facilities, a slight anticorrelation of the median and best 10\% SNRs can be observed.
This can be explained as follows: in all cases the network SNR is dominated by the SNR of the third-generation detector(s) (see Fig.~\ref{fig:gw_horizons}).
For networks with only one third-generation facility, the network SNR is therefore determined by the antenna pattern of the single third-generation facility regardless of its location and orientation.
For networks with two or three third-generation facilities, if these facilities are placed so that their antenna patterns mostly overlap, they will jointly detect events at the antenna pattern maxima with good SNR at the expense of events that are incident close to the antenna pattern minima.
This leads to enhanced SNR for the best 10\% of events and diminished SNR for the median events.
Conversely, if these facilities are placed so that their antenna patterns have little overlap, then the network antenna pattern is more isotropic, leading to better SNR for the median events and worse SNR for the best 10\% of events.

\subsection{Redshift and inclination angle uncertainties for black-hole binaries at $z = 2$}

\begin{figure}[tp]
    \includegraphics[width=\columnwidth]{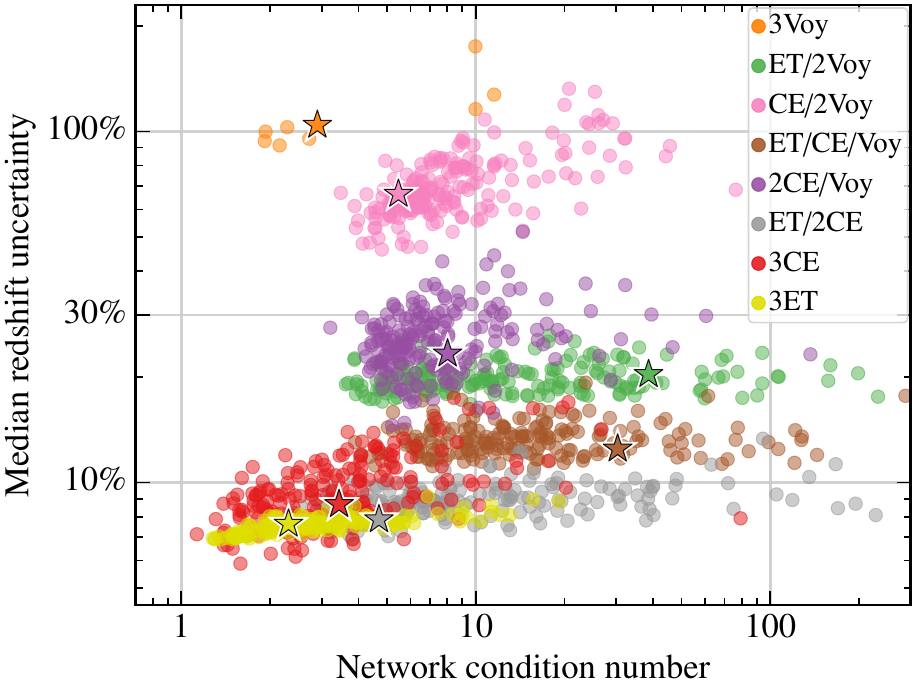}
    \caption{
        Distribution of median fractional redshift uncertainties (90\% confidence) for randomly simulated networks, shown for $z = 2$ and $M_1 = M_2 = 30\,M_\odot$.
        The horizontal axis denotes the condition number (Appendix~\ref{sec:orthogonality}) of the network.
        The HLV point is omitted.
        }
    \label{fig:cbc_med_redshift_bbh}
\end{figure}

Precise measurements of black hole redshifts at $z \gtrsim 1$ can constrain the redshift distribution of black hole binaries, and hence the star formation rate and the delay time from star formation to merger~\cite{Vitale:2018yhm}.
Measurement of the redshift is also critical to determining the source-frame mass of the component black holes, and thus their astrophysical origin.

The distance information in a binary coalescence signal is partially degenerate with the inclination angle $\iota$ (iota) of the binary relative to the line of sight.
Disentangling the two requires good discrimination of the polarization of the incident wave.
To quantify the polarization discrimination ability of a particular network, we can collect the the $N$ detector response tensors~\cite{Dhurandhar:1988my}, each with five independent components, into an $N\times5$ matrix and compute its condition number (appendix~\ref{sec:orthogonality}).
The condition number of the network describes how well the network can reconstruct the parameters of the incoming wave from the detector signals~\cite{Mohanty:2006ha,Rakhmanov:2006qm}.

\begin{figure}[tp]
    \includegraphics[width=\columnwidth]{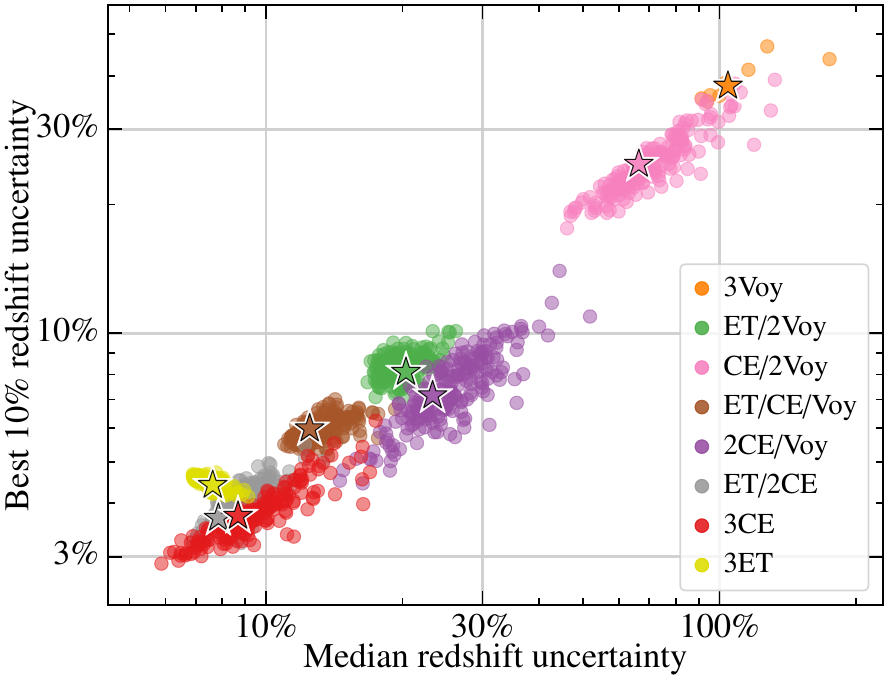}
    \caption{Scatter plot showing median and best 10\% redshift uncertainties (90\% confidence) for $z = 2$ and $M_1 = M_2 = 30\,M_\odot$.
        The HLV point is omitted.}
    \label{fig:cbc_med_better_redshift_bbh}
\end{figure}

\begin{figure}[tp]
    \includegraphics[width=\columnwidth]{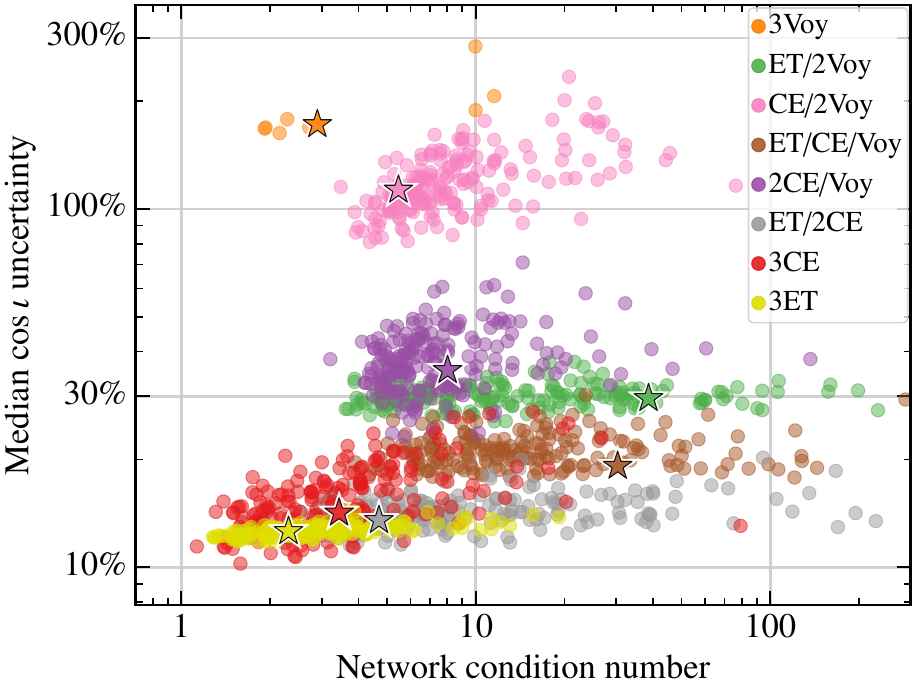}
    \caption{
        Distribution of median fractional uncertainties (90\% confidence) in $\cos\iota$ (the cosine of the inclination angle) for randomly simulated networks, shown for $z = 2$ and $M_1 = M_2 = 30\,M_\odot$.
        The horizontal axis denotes the condition number of the network.
        The HLV point is omitted.
        The median uncertainty $\gtrsim\SI{200}{\percent}$ for the three-Voyager networks and some of the CE--Voyager networks indicates that essentially no information about the inclination angle is recovered from the median event for these networks.
        }
    \label{fig:cbc_med_cosinc_bbh}
\end{figure}

\begin{figure}[tp]
    \includegraphics[width=\columnwidth]{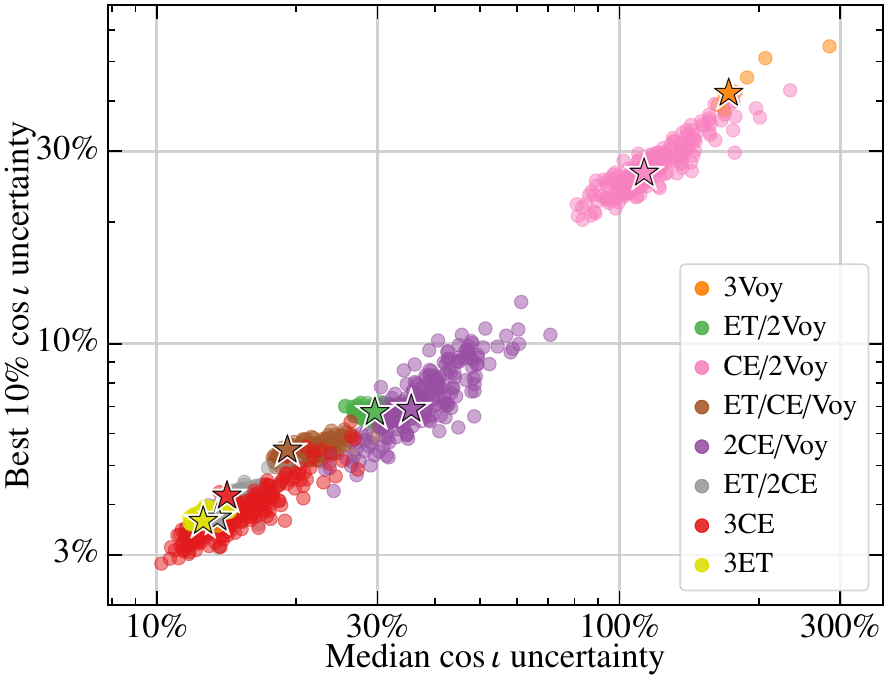}
    \caption{Scatter plot showing median and best 10\% uncertainties (90\% confidence) in $\cos\iota$ for $z = 2$ and $M_1 = M_2 = 30\,M_\odot$.
        The HLV point is omitted.}
    \label{fig:cbc_med_better_cosinc_bbh}
\end{figure}

Fig.~\ref{fig:cbc_med_redshift_bbh} shows the median redshift uncertainties for 30--30\,$M_\odot$ black hole binaries coalescing at $z = 2$, plotted against the network condition number.
Fig.~\ref{fig:cbc_med_better_redshift_bbh} then shows the median versus best 10\% redshift uncertainties for 30--30\,$M_\odot$ black hole binaries coalescing at $z = 2$.
The equivalent plots for the $\cos\iota$ are shown in Figs.~\ref{fig:cbc_med_cosinc_bbh} and~\ref{fig:cbc_med_better_cosinc_bbh}.

The primary conclusion which can be drawn from Figs.~\ref{fig:cbc_med_better_redshift_bbh} and ~\ref{fig:cbc_med_better_cosinc_bbh} is that a network optimized for the best events will also perform well for typical events, which is similar to the conclusion for sky localization from Fig.\ref{fig:cbc_med_better_loc_bns}.
Figs.~\ref{fig:cbc_med_redshift_bbh} and ~\ref{fig:cbc_med_cosinc_bbh}, on the other hand, indicate that while there is some dependence on network condition number, distance and inclination angle uncertainties are not strong functions of detector location and orientation.

\subsection{Signal-to-noise ratios at high frequency}

Gravitational-wave observations at the kilohertz scale can reveal information about nuclear processes from newly merged neutron stars (the so-called ``post-merger'' phase of the waveform)~\cite{Clark:2015zxa,Chatziioannou:2017ixj} and the physics behind core-collapse supernovae~\cite{Gossan:2015xda}.
There is considerable uncertainty in the waveforms produced from these events.
Therefore, we construct a metric consisting of a uniform strain $h(f) = \SI{1e-25}{\per\Hz}$ from \SI{400}{\Hz} to \SI{4}{\kHz} and zero elsewhere; this frequency range encompasses the frequency spectra predicted from multiple neutron-star post-merger models and the expected peak gravitational-wave emission frequencies from core-collapse supernovae.

\begin{figure}[tp]
    \includegraphics[width=\columnwidth]{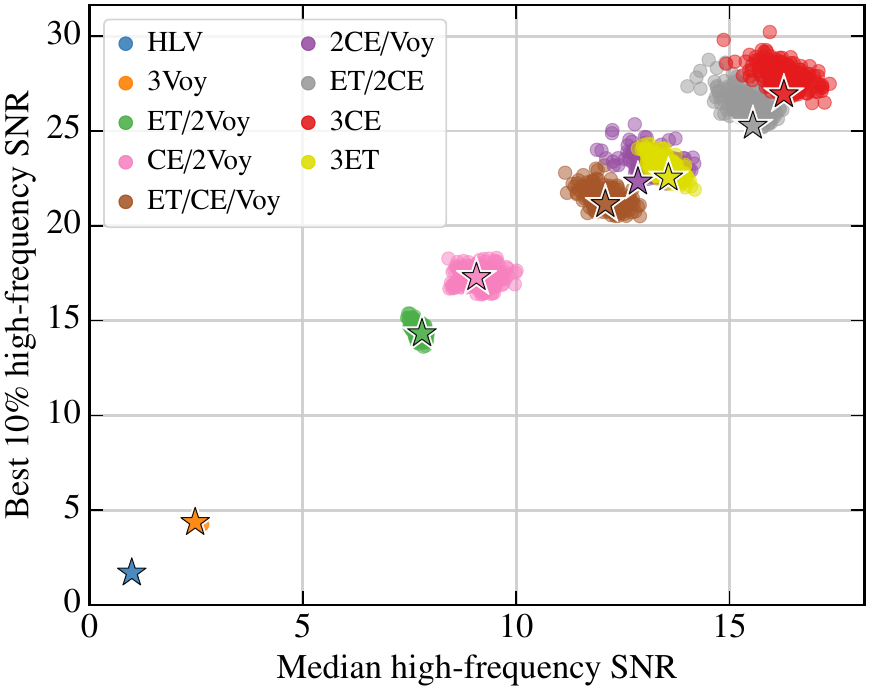}
    \caption{Median and best 10\% high-frequency signal-to-noise ratios for a distribution of 3G networks, assuming a frequency-independent strain $h_0 = \SI{1e-25}{\per\Hz}$.}
    \label{fig:hf_med_better_snr_bns}
\end{figure}

The resulting distributions of signal-to-noise ratios, plotted as median versus best 10\%, is shown in Fig.~\ref{fig:hf_med_better_snr_bns}.
Changing the uniform strain to a frequency-dependent strain $h(f) \propto 1/f^{1/2}$  does not substantively alter the trends shown in Fig.~\ref{fig:hf_med_better_snr_bns}.
The conclusion from this plot is very similar to that of SNR for binary black hole systems in Fig.~\ref{fig:cbc_med_better_snr_bbh}:  while there is an anti-correlation between the best and the median SNR for networks involving 3G detectors, the magnitude of the effect is too small to be a strong driver of network design choices.

\subsection{Stochastic background}

A major goal of gravitational-wave astronomy is to detect a stochastic background from unresolved binary systems, and perhaps from primordial fluctuations in the early universe~\cite{Regimbau:2016ike}.
This background appears as a continuous, broadband signal that is correlated between the detectors in a network.
It can be expressed in terms of a strain power spectral density $S(f)$, but is more often expressed as a strain energy density $\Omega(f) \propto f^3 S(f)$.

\begin{figure}[tp]
    \includegraphics[width=\columnwidth]{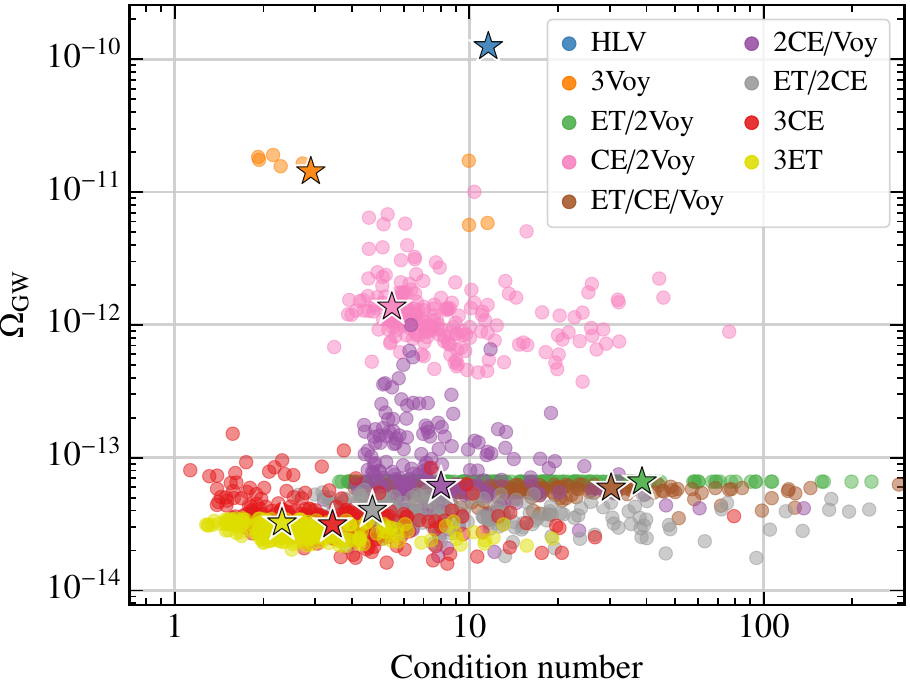}
    \caption{Limits on a white stochastic background, given one year of integration time and a detection threshold SNR of 1.
    The networks are sorted along the horizontal axis by their condition number (Appendix~\ref{sec:orthogonality}.}
    \label{fig:stochastic}
\end{figure}

As a metric, we consider the maximum background that can be resolved with $\text{SNR} = 1$ after one year of integration time~\cite{Allen:1997ad}, assuming the background is white [$\Omega(f) = \Omega_\text{GW} = \text{const.}$] as expected for a primordial background.
Ultimately, the performance of the network in constraining a stochastic background is determined by the network's effective stochastic strain sensitivity, whose power spectral density (PSD) $S_\text{\!eff}(f)$ is given by~\cite{Thrane:2013oya}
\begin{equation}
    \frac{1}{S_\text{\!eff}^2(f)} = \sum_{i=1}^N \sum_{j > i}^N \frac{\Gamma^2_{\!i\!j}(f)}{S_{\!i}(f) S_{\!\!j}(f)},
    \label{eq:stochastic_effective}
\end{equation}
where $\Gamma_{\!i\!j}$ is the overlap reduction function between detectors $i$ and $j$, $S_{\!i}$ and $S_{\!\!j}$ are the detectors' strain PSDs, and $N$ is the total number of detectors in the network.

The resulting limits on $\Omega_\text{GW}$ for different networks are shown in Fig.~\ref{fig:stochastic}, where the networks are again sorted by their condition number.
For the networks comprising single-detector facilities (i.e., no ETs), one should expect an anticorrelation between condition number and the limit on $\Omega_\text{GW}$ since the network's effective noise is minimized when its overlap reduction functions are maximized [Eq.~\eqref{eq:stochastic_effective}], which occurs for colocated and co-oriented detectors.
On the other hand, the networks containing Einstein Telescope facilities already contain triplets of colocated detectors, for which the overlap reduction functions are large regardless of how the facilities are placed.
Thus the stochastic constraining power of the networks with ET facilities show essentially no dependence on the condition number.

Additionally, we find that the network performance for constraining an unresolved white dwarf binary [$\Omega(f) \propto f^{2/3}$] is tightly correlated with the constraining power on $\Omega_\text{GW}$.

\section{Discussion}

In this paper we explore a wide variety of potential future gravitational-wave detector networks and quantify their performance in terms of metrics that determine their scientific potential.
Rather than engaging in a black-box optimization, we connected network performance to physically meaningful quantities, such as network area, which allowed us to draw general conclusions about the importance of various aspects of the network.

First, the performance of a three-facility network is determined primarily by the type of detectors it contains, rather than the location and orientation of the facilities.
This is particularly evident for the signal-to-noise ratio metrics.
For the localization metrics, the network area has a large effect on the performance, but nonetheless this effect is subdominant to the effect of the network composition.

Second, there is a steady progression of scientific potential as more third-generation facilities are added to an existing network of evolved second generation detectors.
This is to say that a single 3G facility will not be sufficient to do all of the science accessible to a full 3G network, nor do we need to have three 3G facilities in order to expand our gravitational-wave potential beyond the second generation.

Lastly, there is not much difference between ranking networks by their median performance or the performance for the best 10\% of events.
In the signal-to-noise ratio metrics one observes a slight anti-correlation, particularly for networks with two or three third-generation facilities, but this is a ${\lesssim}30\%$ effect that is subdominant to sky localization---as seen in Fig.~\ref{fig:cbc_loc_bns}, the largest networks (area $\simeq 1.3r_\oplus^{\,2}$) typically achieve a factor of ${\sim}3$ better BNS localization than networks whose area is similar to the area of the current HLV network ($\simeq 0.3r_\oplus^{\,2}$).

We offer several caveats for the results presented here.
First, the parameter estimation is Fisher-matrix-based, and the results may differ from a full numerical parameter estimation; this is particularly true for the distance and inclination metrics, where priors play an important role.
Second, the systems presented here are localized primarily using timing information; for heavier and higher-redshift systems, amplitude and phase information become important.
This means that the relative orientations of different facilities may become important for these systems.
Third, time-dependent antenna pattern effects were not considered.

With these caveats, we conclude that when designing a three-facility next-generation network, one must think carefully about which detectors should comprise the network.
However, one has a great deal of freedom to choose the locations and orientations of the facilities.

\appendix
\section{Network condition number}
\label{sec:orthogonality}

This appendix describes the network condition number used in the main text~\cite{Mohanty:2006ha,Rakhmanov:2006qm}.
A detector network can be viewed as a linear system that transforms an incoming gravitational-wave strain tensor into a set of detector signals.
The condition number of this linear transformation quantifies how well the strain tensor can be reconstructed from the detector signals, with a smaller condition number indicating better reconstruction.

The astrophysical strain incident on a detector network is described by a symmetric, traceless Cartesian tensor of order 2; its matrix representation is
\begin{equation}
    \mathsf{H} = \begin{pmatrix}
        H_{11}  & H_{12}    & H_{13}    \\
        H_{12}  & H_{22}    & H_{23}    \\
        H_{13}  & H_{23}    & H_{33}
    \end{pmatrix},
\end{equation}
with the additional constraint $H_{11} + H_{22} + H_{33} = 0$.

Correspondingly, the response tensor $\mathsf{D}^{(i)}$ of the $i$th detector to $\mathsf{H}$ is also a symmetric, traceless Cartesian tensor of order 2, defined as the difference of the outer products of the detector's arm vectors $\mathbf{\hat{X}}^{(i)}$ and $\mathbf{\hat{Y}}^{(i)}$:
\begin{equation}
    \mathsf{D}^{(i)} = \frac{1}{2}\left[\mathbf{\hat{X}}^{(i)}\!\otimes\!\mathbf{\hat{X}}^{(i)} - \mathbf{\hat{Y}}^{(i)}\!\otimes\!\mathbf{\hat{Y}}^{(i)}\right].
\end{equation}
The (weighted) signal appearing in the $i$th detector is then given by the contraction of $\mathsf{D}^{(i)}$ and $\mathsf{H}$ into a scalar:
\begin{subequations}
    \begin{align}
        h^{(1)} &= w^{(1)} \mathsf{D}^{(1)} : \mathsf{H} \\
                &\hphantom{!}\vdots        \nonumber \\
        h^{(N)} &= w^{(N)} \mathsf{D}^{(N)} : \mathsf{H}.
    \end{align}
\label{eq:SNRstrains}
\end{subequations}
The weights $\{w^{(i)}\}$ account for the fact that in a heterogeneous network, some detectors respond to the incident strain tensor with greater fidelity than others.
In this work we choose to weight the signals by the detector noise performance at \SI{100}{\Hz}: $w^{(i)} = 1/\sqrt{S^{(i)}(\SI{100}{\Hz})}$, where $S^{(i)}(f)$ is the power spectral density of the detector's strain noise.

A symmetric, traceless, order-2 tensor in $\mathbb{R}^3$ has five independent components,
meaning that both $\mathsf{H}$ and the detector tensors can be written in terms of five basis vectors $\mathsf{E}_1,\ldots,\mathsf{E}_5$~\cite{Dhurandhar:1988my}~\footnote{Note, however, that a strain tensor produced by a single source on the sky in fact has only four independent components---the sky locations $\theta$ and $\phi$, and the plus- and cross-polarization amplitudes $h_+$ and $h_\times$~\cite{Guersel:1989th}.
Similarly, a detector with two equal-length arms also only has four independent components, two angles $\theta_\mathrm{X}$ and $\phi_\mathrm{X}$ to describe the orientation of the X arm and $\theta_\mathrm{Y}$ and $\phi_\mathrm{Y}$ for the Y arm.}.
The Cartesian matrix representation of one such (orthonormal) basis is
\begin{subequations}
    \begin{align}
        \mathsf{E}_1 &= \frac{1}{\sqrt{2}}
            \begin{pmatrix}
                1 &   0 &   0 \\
                0 &  -1 &   0 \\
                0 &   0 &   0
            \end{pmatrix}   \\
        \mathsf{E}_2 &= \sqrt{\frac{2}{3}}
            \begin{pmatrix}
              \tfrac{1}{2} &    0 &   0 \\
                0 &  \tfrac{1}{2} &   0 \\
                0 &    0 &  -1
            \end{pmatrix}   \\
        \mathsf{E}_3 &= \frac{1}{\sqrt{2}}
            \begin{pmatrix}
                0 &   1 &   0 \\
                1 &   0 &   0 \\
                0 &   0 &   0
            \end{pmatrix}   \\
        \mathsf{E}_4 &= \frac{1}{\sqrt{2}}
            \begin{pmatrix}
                0 &   0 &   1 \\
                0 &   0 &   0 \\
                1 &   0 &   0
            \end{pmatrix}   \\
        \mathsf{E}_5 &= \frac{1}{\sqrt{2}}
            \begin{pmatrix}
                0 &   0 &   0 \\
                0 &   0 &   1 \\
                0 &   1 &   0
            \end{pmatrix}.
    \end{align}
\end{subequations}
With this representation, the system of equations \eqref{eq:SNRstrains} can instead be written as a matrix equation

\begin{equation}
    h^{(i)} = M^{(i\hspace{-0.2ex}j)} H^{(j)},
\end{equation}
where $H^{(j)} = \mathsf{H}:\mathsf{E}^{(j)}$ are the elements of a 5-element column vector, and $M^{(i\hspace{-0.2ex}j)} = w^{(i)}\mathsf{D}^{(i)}:\mathsf{E}^{(j)}$ are the elements of an $N\times5$ matrix $\mathcal{M}$.

One can then define the condition number of the network as
\begin{equation}
    \kappa(\mathcal{M}) = \sigma_\text{max}(\mathcal{M}) \times \sigma_\text{max}(\mathcal{M}^+),
\end{equation}
where $\sigma_\text{max}(\cdot)$ is the maximum singular value, and $\mathcal{M}^+$ is the pseudoinverse of $\mathcal{M}$.

\begin{acknowledgments}
    EDH is supported by the MathWorks, Inc.
    NSF award PHY-1836814 supports third-generation gravitational-wave detector network research in the United States.
    The authors thank R. Eisenstein for careful reading of the manuscript, as well as S. Vitale, T. Callister and M. Isi for helpful discussions.
    This article has been assigned the document number LIGO-P1800304.
\end{acknowledgments}

\bibliographystyle{apsrev4-2}
\bibliography{3g-metrics-paper}
\end{document}